\documentstyle[aps,epsf,floats]{revtex}

\ifx\epsffile\undefined
\def\insertfig#1{}% null macro
\else
\def\insertfig#1{{\baselineskip=4pt
\centerline{\epsfxsize=\hsize\epsffile{#1}}}}\fi

\begin{document}

\twocolumn[       %typeset the title and abstract in one column
{\tighten
\preprint{\vbox{
\hbox{DOE/ER/40427-28-N96}
\hbox{nucl-th/9608036}
}}

%\draft
\title{The role of chiral symmetry in two-pion exchange nuclear 
potential\footnotemark}
\author{C.A. da Rocha\footnotemark}
\address{University of Washington, Dept. 
of Physics, Box 351560, Seattle, Washington, 98195-1560}
\author{M.R. Robilotta\footnotemark} 
\address{University of Washington, Dept. 
of Physics, Box 351560, Seattle, Washington, 98195-1560 \\
and \\
Instituto de F\'{\i}sica, Universidade de S\~ao Paulo,
C.P. 20516, 01452-990, S\~ao Paulo, SP, Brazil }
\author{J.L. Ballot\footnotemark}
\address{Division de Physique Th\'eorique,
Institut de Physique Nucl\'eaire, F91406 Orsay C\'edex, France}
\bigskip
\date{August 96}
\maketitle
\widetext
\vskip-2.8in
\rightline{\vbox{
\hbox{DOE/ER/40427-28-N96}
}}
\vskip2.45in
\begin{abstract}
We evaluate the two pion exchange contribution to the
nucleon-nucleon potential in configuration space using a {\em minimal
chiral model} containing only pions and nucleons.
We argue that this model has nowadays a rather firm conceptual basis,
which entitles it to become a standard ingredient of any modern potential.
The main features of this model is that the scalar-isoscalar
component of the interaction is relatively small, due to cancellations
between large terms, and fails to reproduce the intermediate range
attraction in the central channel. We show that chiral symmetry is
the responsible for these large cancellations in the
two-pion exchange nucleon-nucleon interaction, which are similar to those
occuring in free pion-nucleon scattering.
Another feature of the model is that these results do
not depend on how chiral symmetry is implemented.
\end{abstract}

}% end the tighten

] % end twocolumn format
\narrowtext

\footnotetext{${}^*$Complete version of the contribution presented at
the $14^{\text{th}}$ International Conference on Particles and Nuclei 
(PANIC 96), Williamsburg, VA, 22-28 May 1996. }
\footnotetext{${}^\dagger$Fellow from CNPq, Brazilian Agency.
Eletronic address: carocha@phys.washington.edu}
\footnotetext{${}^\ddagger$Eletronic address: robilotta@if.usp.br}
\footnotetext{${}^{\text{\S}}$Unit\'e de Recherche des Universit\'es Paris 6 
et Paris 11 associ\'ee au CNRS. Eletronic address: ballot@frcpn11.in2p3.fr}

%%%%%%%%%%%%%%%%%%%%%%%% SECTION 1
\vskip-0.2in
\section{INTRODUCTION}
\label{sec:introduction}

The long range part of the nucleon-nucleon $(NN)$ interaction is ascribed
to the exchange of one pion (OPEP) and is very well known.  The medium range
region, on the other hand, is much more controversial, since
in the literature one finds various competing theoretical approaches, as 
dispersion relations, field theory or just phenomenology. In all cases,
the medium range interaction is associated with the exchange of two pions
(TPEP), whereas there is a wide variation in the way short distance effects 
are treated. In order to reproduce the data, parameters are used which either
reflect knowledge about other physical processes or are adjusted ad hoc. 
The differences in all these approaches are enhanced in the isospin-symmetric 
nuclear matter, where the OPEP vanishes and one must know the asymptotic
behavior of the TPEP. 

The role of the TPEP in the framework of chiral symmetry has recently attracted
considerable attention, especially as far as the restricted pion-nucleon sector
is concerned~\cite{Ordo92,Cele92,Fria94,Rocha94,Birse94,Ordo94}. It is well 
known that this process is closely related to the
pion-nucleon scattering amplitude \cite{Rocha94,RR95}.
 
Chiral symmetry at hadron level may be implemented by means of either linear or
non-linear Lagrangians. The fact that no serious candidate has been found for
the $\sigma$ meson favours the latter type of approach. There are two forms of
non-linear Lagrangians which are especially suited for the $\pi$N system. One
of them is based on a pseudoscalar (PS) $\pi$N coupling supplemented by a
scalar (S) interaction, equivalent to the exchange of
an infinitely massive $\sigma$ meson, and
denoted as $PS+S$ scheme (Eq.~\ref{Eq.1}). The other one employs a pseudovector
(PV) $\pi$N interaction and a vector (V) term, which could represent the
exchange of an infinitely massive $\rho$ meson, constituting the $PV+V$ scheme
(Eq.~\ref{Eq.2}).
\begin{eqnarray}
{\cal L}_{PS+S} &=& \cdots - g \overline{\psi}\,
  \left(\sqrt{f_{\pi}^2 - \phi^2} + i \bbox{\tau} \cdot \bbox{\phi}
  \gamma_5\right) \psi\;, \label{Eq.1} \\ [0.3cm]
{\cal L}_{PV+V} &=& \cdots - \frac{1}{4 f_{\pi}^2}\,
   \overline{\psi}\,\gamma^\mu\,\bbox{\tau}\,\psi\cdot
   \bbox{\phi} \times \partial_\mu\,\bbox{\phi} \nonumber \\ [0.3cm]
   &+&\frac{g_A}{2f_\pi}\,\overline{\psi}\,\gamma^{\mu}
   \gamma_5\, \bbox{\tau} \,\psi\cdot\partial_\mu\,
  \bbox{\phi} + \cdots\;.
\label{Eq.2}
\end{eqnarray}

\noindent
Both approaches yield the very same amplitude for the $\pi$N scattering 
for the isospin-symmetric amplitude. The fact that physical results should be
independent of the representation used to implement chiral symmetry was
discussed in very general terms by Coleman, Wess, and Zumino \cite{CWZ69}.

As far as nucleon-nucleon scattering is concerned, the two-pion exchange
amplitude up to order $1/f_\pi^4$ is given by five diagrams, usually named box
($\Box$), crossed box \hfill
 
\begin{figure}
\epsfxsize=15.0cm
\centerline{\epsffile{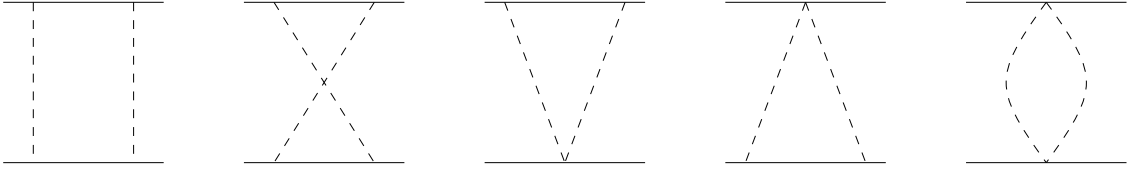}}
\vspace{0.7cm}
\caption{Loop diagrams for the two pion exchange NN potential calculated in
the minimal chiral model.}
\label{Fig.1}
\end{figure}

\onecolumn 
\noindent
($\Join$), triangle ($\triangle+\nabla$) and bubble
($(\!)$), given in Fig.~\ref{Fig.1}, which constitute the {\sl  minimal chiral
model} or the TPEP.

%%%%%%%%%%%%%%%%%%%%%%%%% FIGURE 1

\noindent The first two diagrams contain only 
nucleon propagators and are independent of chiral symmetry, whereas the 
triangles and the bubble involve the scalar interaction and hence are due to 
the symmetry. When one considers the potential instead of the amplitude, the 
iterated OPEP has to be subtracted from the box diagram.

\section{Parametrization of the TPEP}

Our calculation of the TPEP is based on the Blankenbecler-Sugar 
reduction of the Bethe-Salpeter equation. Its dynamical 
content is associated with the five diagrams displayed in Fig.~\ref{Fig.1}.
Therefore we label the corresponding individual contributions by 
${(\!)}\;,\triangle\;,{\Join}\mbox{ and }\Box$, where the last one also 
includes the subtraction of the iterated OPEP. 
It has the general form

\begin{eqnarray}
V(r)&=&\left[\left(V_{(\!)}^C+V_\triangle^C\right)+\hat{O}_{LS}
\left(V_{(\!)}^{LS}+V_\triangle^{LS}\right)\right]
+\left(3+2 \bbox{\tau}^{(1)}\cdot \bbox{\tau}^{(2)}\right) 
\left[V_{\Join}^C+\hat{O}_{SS}V_{\Join}^{SS}+\hat{O}_{LS}V_{\Join}^{LS}
+\hat{O}_T V_{\Join}^T\right] \nonumber \\ [0.3cm]
&+&\left(3 - 2\bbox{\tau}^{(1)}\cdot\bbox{\tau}^{(2)}\right)
\left[V_\Box^C+\hat{O}_{SS}V_\Box^{SS}+\hat{O}_{LS}V_\Box^{LS}+\hat{O}_T
V_\Box^T\right]
\label{Eq9}
\end{eqnarray}%(9)

\noindent
where the spin operators are given by 
$\hat{O}_{SS}=\bbox{\sigma}^{(1)}\cdot\bbox{\sigma}^{(2)}$, 
$\hat{O}_{LS}={\bf L}\cdot\case{1}/{2}\{\bbox{\sigma}^{(1)}+
\bbox{\sigma}^{(2)}\}$, and $\hat{O}_T=3\bbox{\sigma}^{(1)}\cdot\hat{\bf r}
\;\;\bbox{\sigma}^{(2)}\cdot\hat{\bf r}-\bbox{\sigma}^{(1)}\cdot
\bbox{\sigma}^{(2)}$, whereas $\bbox{\sigma}^{(i)}$ and $\bbox{\tau}^{(i)}$
represent spin and isospin matrices for nucleon $(i)$.

In the case of the bubble diagram, the leading contribution to the asymptotic 
potential can be calculated analytically~\cite{Rob95}; its result
sets the pattern for the parametrization of the other components of the force.

Our numerical expressions represent the various components of the potential in 
MeV, and are given in terms of the adimensional variable $x\equiv\mu r$, where
$\mu$ is the pion mass. We keep the axial 
$\pi$N coupling constant $g_A$ as a free 
parameter  and adopt the values $\mu=137.29$ MeV and $f_\pi=93$ MeV for the 
pion mass and the pion decay constant respectively.

In general, the parametrized expressions reproduce quite well the numerical 
results for the complete theoretical calculations, given in
Ref.~\cite{Rocha94}, except for a few cases and regions where the 
discrepancies become of the order of 0.25\%. Our results are listed below.

\subsection{Central Potential}

The profile function for the central potential has the following common 
multiplicative expression
\begin{equation}
F_c(x)=\left(\frac{g_A\;\mu}{2f_\pi}\right)^4\;\frac{e^{-2x}}{x^2\sqrt{x}}\;.
\end{equation}

The parametrization of each diagram gives:
\begin{eqnarray}
V^C_{(\!)}(x)&=&F_c(x)\left\{-275.364-\frac{51.0923}{x}+\frac{6.54068}{x^2}-
\frac{1.26190}{x^3}+\frac{0.130706}{x^4}\right\}
\label{vc01} \\ [0.3cm]
V^C_{\triangle}(x)&=&F_c(x)\left\{343.558-\frac{14.0446}{x}
+\left(135.249+14.6514\cdot x+6.43825\cdot x^2\right)\cdot 
e^{-0.397835\cdot x}\right\}
\label{vc02}
\\ [0.3cm]
V^C_{\Join}(x)&=&\frac{V^C_{\triangle}(x)}{12}+F_c(x)\left\{
-\frac{265.304}{\sqrt{x}}+\frac{518.531}{x}
-\frac{577.210}{x\sqrt{x}}+\frac{378.004}{x^2}-
\frac{133.374}{x^2\sqrt{x}}+\frac{19.5061}{x^3}\right\}
\label{vc12}
\\ [0.3cm]
V^C_{\Box}(x)&=&F_c(x)\left\{-25.9987+\frac{8.33777}{x}-\frac{0.870724}{x^2}
\right\}\cdot e^{-[0.101214\cdot x+0.00123687\cdot x^2]}
\label{vc09}
\end{eqnarray}

\subsection{Spin-spin potential}

The multiplicative factor for the spin-spin potential is the same as that of 
the central potential, and receives contributions from the box and crossed 
diagrams only, which are given by
\begin{eqnarray}
V^{SS}_{\Join}(x)&=&F_c(x)\left\{0.408084+\frac{1.05042}{x}+\frac{0.421043}
{x^2}-\frac{0.0284309}{x^3}-0.215829\;\cdot e^{-1.2344\cdot x}\right\}
\label{vss12} \\ [0.3cm]
V^{SS}_{\Box}(x)&=&F_c(x)\left\{0.399845+\frac{1.07191}{x}+\frac{0.216302}{x^2}
-\frac{0.0306271}{x^3}+0.0371333\,x\cdot e^{-0.16808\cdot x}\right\}
\label{vss09}
\end{eqnarray}

\subsection{Spin-orbit Potential}

The spin-orbit multiplicative function is 
\begin{equation}
F_{LS}(x)=\left(\frac{g_A\;\mu}{2f_\pi}\right)^4\left(1+\frac{1}{2x}\right)
\;\frac{e^{-2 x}}{x^3\sqrt{x}}\;,
\end{equation}
\noindent
and individual contributions are:
\begin{eqnarray}
V^{LS}_{(\!)}(x)&=&F_{LS}(x)\left\{-5.88744-\frac{5.51078}{x}+\frac{0.994157}
{x^2}-\frac{0.217562}{x^3}+\frac{0.0336067}{x^4}-\frac{0.00244620}{x^5}
\right\}
\label{vls01} \\ [0.3cm]
V^{LS}_{\triangle}(x)&=&F_{LS}(x)\left\{7.34548+\frac{2.15233}{x}-
\frac{0.381025}{x^2}
+\left(7.48798-0.448484\cdot x+0.391431\cdot x^2\right)
\cdot e^{-0.419984\cdot x}\right\}
\label{vls02} \\ [0.3cm]
V^{LS}_{\Join}(x)&=&-\frac{V^{LS}_{\triangle}(x)}{4}+F_{LS}(x)\left\{
\frac{5.67235}{\sqrt{x}}-\frac{10.6417}{x}
+\frac{14.7389}{x\sqrt{x}}-\frac{12.6506}{x^2}+\frac{6.27374}
{x^2\sqrt{x}}-\frac{1.66637}{x^3}+\frac{0.184264}{x^3\sqrt{x}}\right\}
\label{vls12} \\ [0.3cm]
V^{LS}_{\Box}(x)&=&F_{LS}(x)\left\{-1.19527-\frac{1.58089}{x}+\frac{0.319790}
{x^2}-\frac{0.0196461}{x^3}\right\}\cdot\left[1-0.0160259\cdot x\right]^{-1}
\label{vls09}
\end{eqnarray}

\subsection{Tensor Potential}

The common factor for the tensor potential is
\begin{equation}
F_T(x)=\left(\frac{g_A\;\mu}{2f_\pi}\right)^4\left(1+\frac{3}{2x}+
\frac{3}{4x^2}\right)\;\frac{e^{-2 x}}{x^2\sqrt{x}}\;.
\end{equation}
\noindent
It receives contributions from the box and crossed diagrams only, which have
the form
\begin{eqnarray}
V^T_{\Join}(x)&=&F_T(x)\left\{-0.204041-\frac{0.510720}{x}+\frac{0.0597556}
{x^2}
+\left(1.32932-0.939553\cdot x+0.706050\cdot x^2\right)
\cdot e^{-2.29686\cdot x}\right\}\label{vt12} \\ [0.3cm]
V^T_{\Box}(x)&=&F_T(x)\left\{-0.246349-\frac{0.521123}{x}+\frac{0.352463}{x^2}
-\frac{0.135028}{x^3}+\frac{0.0303012}{x^4}-\frac{0.00297840}{x^5}
\right\}
\label{vt09}
\end{eqnarray}

\twocolumn 

\section{Results}
 
\begin{figure}
%\vspace{0.5cm}
\epsfxsize=7.0cm
\epsfysize=8.0cm
\centerline{\epsffile{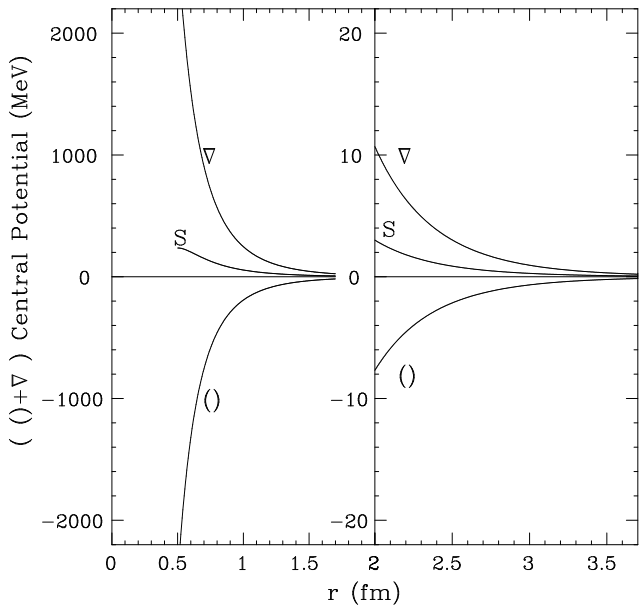}}
\vspace{0.5cm}
\caption{Profile functions for the bubble ($(\!)$) and triangle
($\nabla$) scalar-isoscalar potentials and for their  sum 
(S), showing a strong cancellation between 
these two contributions. The graph at right is an amplification.}
\label{Fig.2}
\end{figure}

Using the minimal chiral model potential described in last section 
we evaluated the scalar-isoscalar component of the TPEP and the phase 
shifts for some singlet waves. We adopted $g_A=1.33$~\cite{Ordo94}.
It is possible to 
notice two important cancellations within the scalar-isoscalar sector of the 
$\pi\pi$E-NNP. The first of them happens between the triangle and bubble 
contribution, as shown in Fig.~\ref{Fig.2}. 

%%%%%%%%%%%%%%%%%%%%%%%%%%%%%% FIGURE 2

\begin{figure}
%\vspace{0.5cm}
\epsfxsize=7.0cm
\epsfysize=8.0cm
\centerline{\epsffile{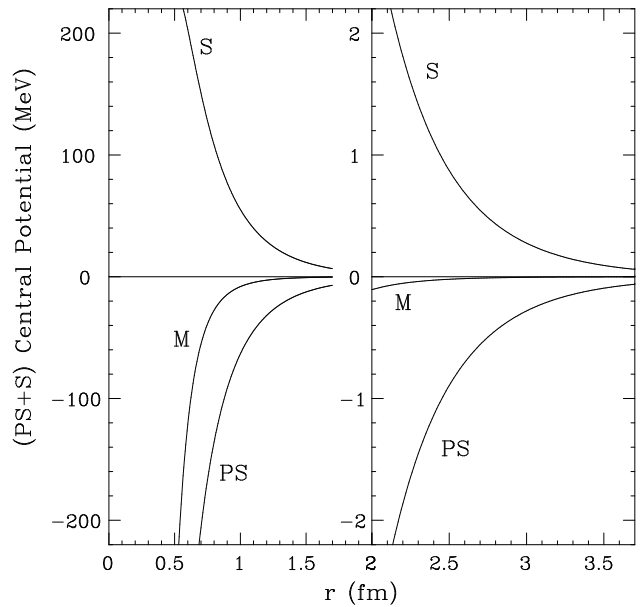}}
\vspace{0.5cm}
\caption{Profile functions for the chiral (S) and pseudoscalar
(PS) scalar-isoscalar potentials  and for their sum 
(M), showing another strong cancellation between 
these two contributions. The graph at right is an amplification.}
\label{Fig.3}
\end{figure}

The other one occurs when the 
remainder from the previous cancellation (S) is added to the sum of the box 
and crossed box diagrams (PS). In this last case, the direct inspection of the 
profile functions for the potential, given in Fig~\ref{Fig.3}, provides 
just a rough estimate of the importance of the cancellation, 
since the iterated OPEP is not included there. 

%%%%%%%%%%%%%%%%%%%%%%%%%%%%%% FIGURE 3

The second cancellation can be better studied in the $NN$ scattering problem, 
since the amplitudes obtained by solving a dynamical equation include 
automatically the iterated OPEP and depend  very little on the way the
potential is defined. In our case, we 
are interested in exhibiting the effects associated with chiral symmetry in the 
scalar-isoscalar two-pion exchange channel. Therefore we concentrate our study 
on singlet channels, where the strong effects associated with tensor OPEP 
interactions are not present. A suitable choice of observables also allow a 
separation of the dynamical effects according to their range. This is 
particularly useful in this problem, since it  is well known that the use of
the Chiral Perturbation Theory (CHPT) is associated with the  inclusion 
of undetermined counterterms in the Lagrangian, involving higher orders of the
relevant momenta \cite{Ordo92,Ordo94,BKU95}. 
However, in configuration space, these
counterterms become delta functions which affect just the origin and hence are
effective only for waves with low orbital angular momentum. In our derivation
of the NN potential we used a Lagrangian which did not contain these contact
terms and hence it is suited for medium and long distances. Thus, in order to
avoid these undetermined short range effects, we consider only the 
$^1D_2,\;^1G_4,\;^1F_3$, and $^1H_5$ waves.
For each channel, we decompose the full $NN$ potential $V$ as 
\begin{equation}
V=U_\pi+U_{PS}+U_S+U_C\;,
\label{Eq.2m}
\end{equation}
\noindent
where $U_\pi$ is the OPEP, $U_C$ represents the short ranged core 
contributions, $U_{PS}$ is due to the box and crossed box diagrams whereas 
$U_S$ is associated with the chiral triangle and bubble interactions. Using 
the variable phase method, it is possible to write the phase shift for 
angular momentum $\ell$ as \cite{Cal63,BR94}
\begin{equation}
\delta_\ell=-\frac{m}{k}\int_0^\infty dr\;V\;P_\ell^2\;.
\end{equation}
\noindent
In this expression, the structure function $P_\ell$ is given by 
\begin{equation}
P_\ell=\hat{\jmath}_\ell\,\cos{D_\ell} -\hat{n}_\ell\,\sin{D_\ell}\;,
\end{equation}
\noindent 
where $\hat{\jmath}_\ell$ and $\hat{n}_\ell$ are the usual Bessel and Neumann 
functions multiplied by their arguments and $D_\ell$ is the variable phase. 
Using the decomposition of the potential given in Eq.~\ref{Eq.2}, one writes 
the perturbative result
\begin{eqnarray}
\delta_\ell&=&-\frac{m}{k}\int_0^\infty dr\left\{U_\pi\;\hat{\jmath}^2_\ell 
\right.\nonumber \\ [0.2cm]
&+&\left.\left[U_\pi\left
(P^2_\ell-\hat{\jmath}^2_\ell\right)+U_{PS}P^2_\ell+U_SP^2_\ell\right]
+U_CP^2_\ell\right\}\nonumber \\  [0.5cm]
&\equiv&\delta_\ell)_{\pi L}+\left[\delta_\ell)_{\pi I}+\delta_\ell)_{PS}+
\delta_\ell)_{S}\right]+\delta_\ell)_C\;.
\label{Eq.5}
\end{eqnarray}
\noindent
In this expression, the first term represents the perturbative long range OPEP
($\pi$L), the second the iterated OPEP ($\pi$I), the third the part due to 
the box and crossed-box diagrams (PS), the fourth the contribution from chiral 
symmetry (S). The last one is due to the core and vanishes for waves with 
$\ell\neq0$. 

\begin{figure}
%\vspace{0.5cm}
\epsfxsize=6.5cm
\epsfysize=7.5cm
\centerline{\epsffile{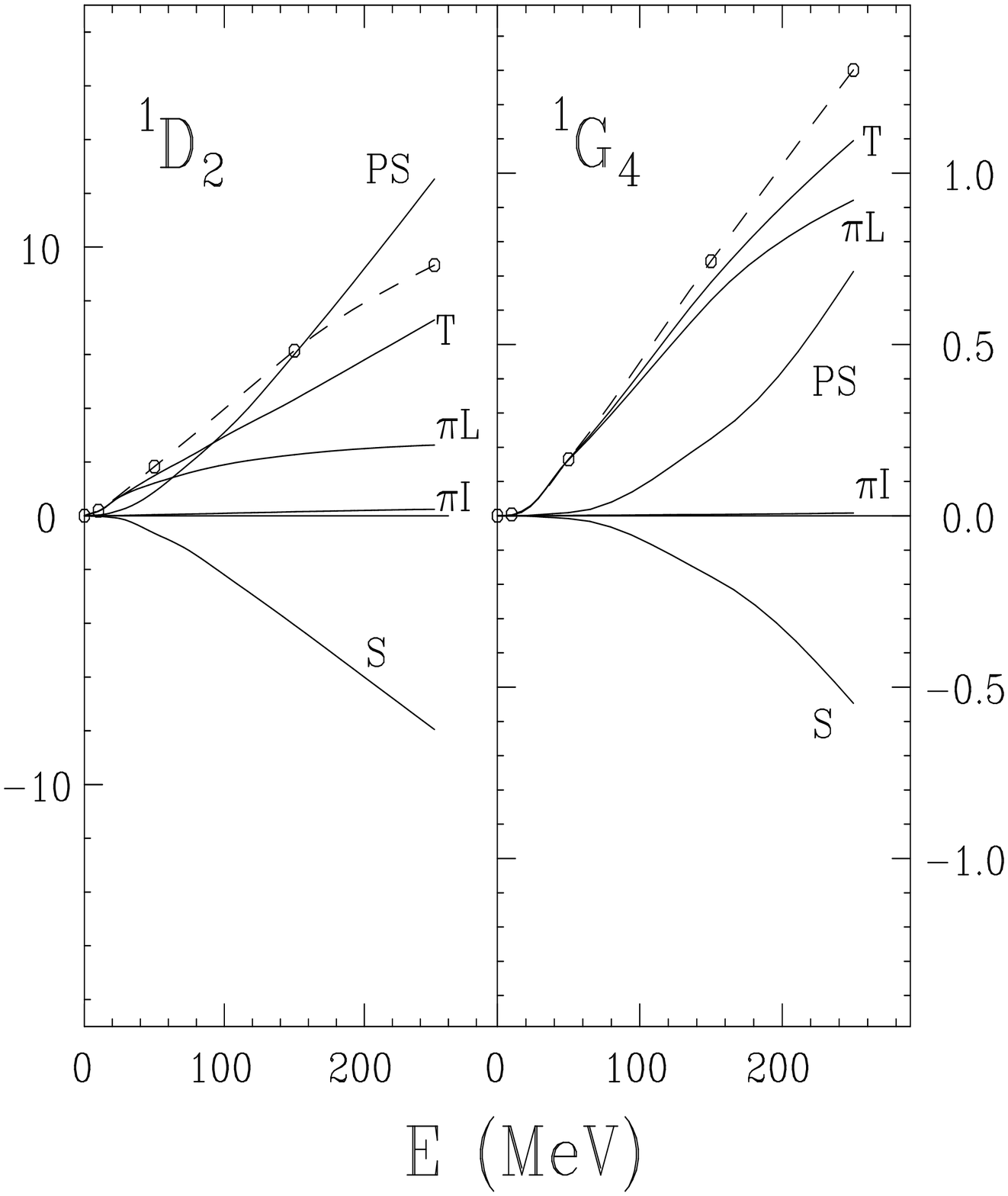}}
\vspace{0.5cm}
\caption{Contributions for the long-OPEP ($\pi$L), iterated OPEP ($\pi$I), 
pseudoscalar (PS) and chiral (S) terms of the potential to the phase shifts 
for $^1D_2$ and $^1G_4$ waves. The total phase shifts are indicated by (T).
The experimental result is given by the dashed line.}
\label{Fig.4}
\end{figure}

In Fig.~\ref{Fig.4} we show the partial contributions of the singlet even 
waves $^1D_2$ and $^1G_4$ to the phase shifts as functions of energy.

%%%%%%%%%%%%%%%%%%%%%%%%%%%%%% FIGURE 4

\begin{figure}
%\vspace{0.5cm}
\epsfxsize=6.5cm
\epsfysize=7.5cm
\centerline{\epsffile{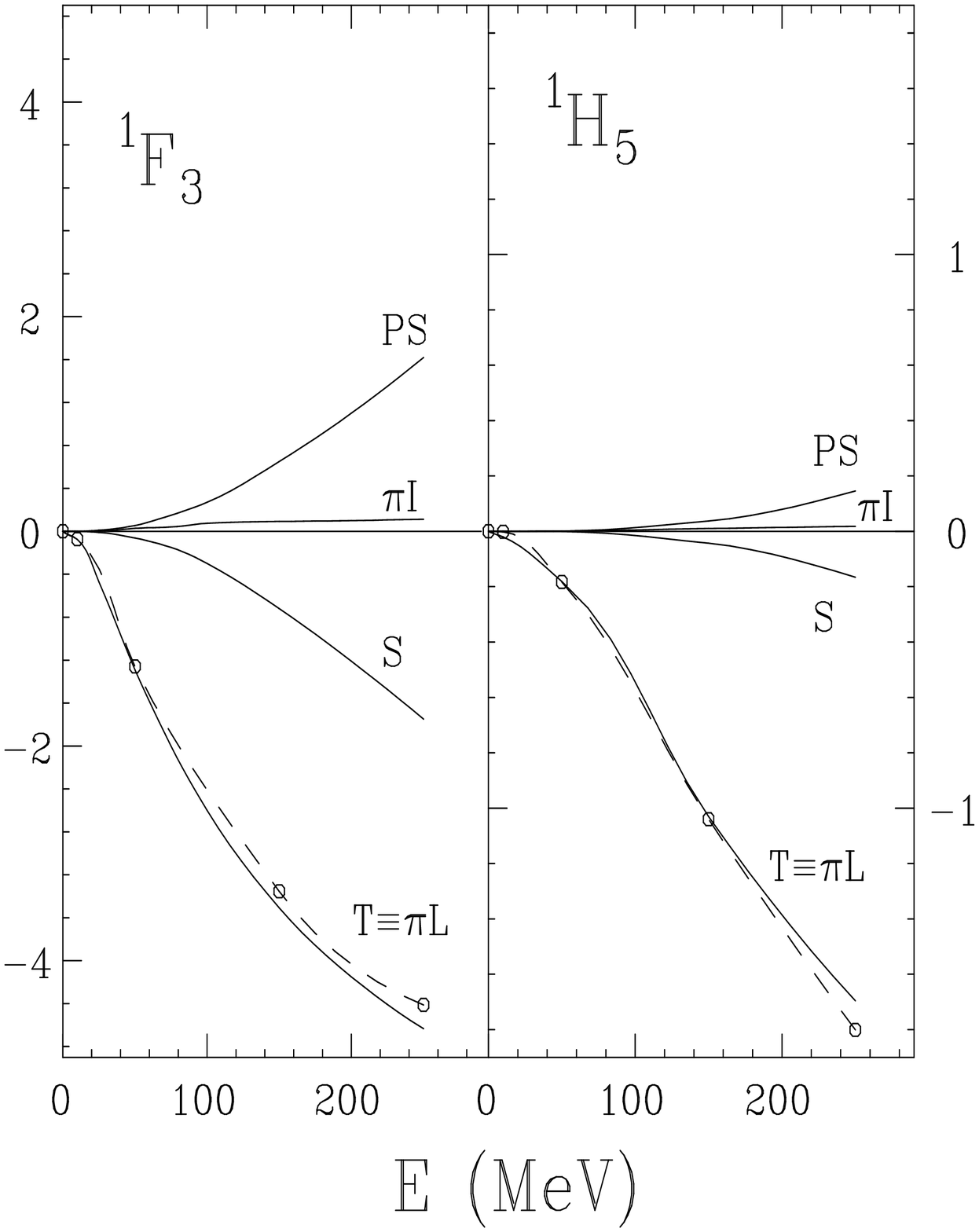}}
\vspace{0.5cm}
\caption{Contributions for the long-OPEP ($\pi$L), iterated OPEP ($\pi$I), 
pseudoscalar (PS) and chiral (S) terms of the potential to the phase shifts 
for $^1F_3$ and $^1H_5$ waves. The total phase shifts are indicated by (T).
The experimental result is given by the dashed line.}
\label{Fig.5}
\end{figure}

\noindent
Observing Fig.~\ref{Fig.4}, one sees that the minimal chiral model with
a core added fails to reproduce the energy dependence of $^1D_2$ wave 
and is resonable  for the $^1G_4$ wave. This not-so-well agreement was
already expected since the minimal chiral model does not include the 
complete dynamics of the $\pi N$ scattering. Mesons and baryons resonances
like $\rho$ and $\Delta$ are very important for higher energies and lower
orbital angular momenta $\ell$. As one goes to higher values of $\ell$, 
the role of the one and two pion exchange becomes more important. This can 
be seen in Fig.~\ref{Fig.5}, where we show the partial contributions of the
singlet odd waves $^1F_3$ and $^1H_5$ to the phase shifts as functions of
energy. 

%%%%%%%%%%%%%%%%%%%%%%%%%%%%%% FIGURE 5

\noindent
A remarkable feature of these two waves is that the net result is given just 
by the long-OPEP contribution, meaning that the medium range contributions 
cancel entirely. In these channels, the iterated OPEP is noticeable and the
relationship $\delta_\ell)_{\pi I}+\delta_\ell)_{PS}=-\delta_\ell)_S$ holds.
This is an important feature of the chiral symmetry and explains why the first 
theoretical models for the TPEP in the 50's, which do not have the $S$ term, 
spoiled the good agreement achieved by the OPEP alone for phase shifts with
large $\ell$.
Moreover, it is possible to see another important features of the TPEP, namely  
the iterated OPEP contribution is comparatively small, indicating that 
ambiguities in the definition of the potential do not have numerical
significance. 

Our results show that chiral symmetry, in the restricted pion-nucleon sector,
is responsible for large cancellations in the two-pion exchange interaction. 
This process is therefore similar to threshold pion-nucleon or 
pion-deuteron~\cite{RW78} scattering amplitudes, 
where the main role of the symmetry is to
set the scale to the problem to be small. 

\section{Perspectives}

A shortcoming of the minimal chiral model is that it fails to reproduce 
experimental information in the case of the intermediate $\pi N$ amplitude 
that enters the TPEP. In order to overcome this difficulty, one may extend the 
approach so as to encompass other degrees of freedom. This possibility
was recently considered by the introduction of the $\Delta$ 
resonance~\cite{Ordo94}, which improved considerably the predictive power of
the method. Another way to achieve this goal is to introduce the empirical
information that \hfill

\begin{figure}
\epsfxsize=16.0cm
\centerline{\epsffile{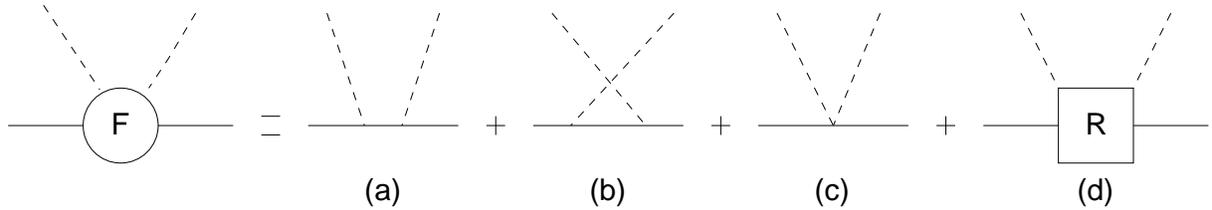}}
\vspace{0.5cm}
\caption{Diagrams associated with the $\pi N$ amplitude. (a), (b), and (c) 
represent the minimal chiral model, whereas  (d) corresponds to the net effect
of the HJS coeficients and is denoted by $R$, for ``rest'' .}
\label{Fig.6}
\end{figure}

\onecolumn
\noindent is missing in the intermediate $\pi N$ amplitude in a model
independent way, with the help of the H\"ohler, Jacob and
Strauss~\cite{HJS72,KP80,HOH83} subthreshold coeficient, as proposed by 
Robilotta~\cite{Rob95}.
The detailed knowledge of the $\pi N$ amplitude provided by these coeficients 
allows a precise determination of the missing part of the TPEP, as illustrated
in Fig.~\ref{Fig.6}.
This approach is very similar to in spirit to that followed long ago by 
Tarrach and Ericson~\cite{TE78}, who explored in detail the analogy between 
the TPEP and Van der Waals force.

%%%%%%%%%%%%%%%%%%%%%%%%% FIGURE 6

\noindent
A detailed calculation for the leading term in this approach was already 
done~\cite{Rob95}, and a complete calculation will be presented 
soon~\cite{Rocha96}.

\acknowledgements

We thank Prof. U. van Kolck for helpful conversations about Chiral 
Perturbation Theory and the $NN$ force. This work was partially
supported by U.S. Department of Energy.
The work of C.A. da Rocha was supported by CNPq Brazilian Agency.


\begin{references}
\bibitem{Ordo92} C. Ord\'{o}\~nez and U. Van Kolck, {\sl Phys. Lett. B}
{\bf 291}, 459 (1992).

\bibitem{Cele92} L.S. Celenza, A. Pantziris and C.M. Shakin, {\sl Phys.
Rev. C} {\bf 46}, 2213 (1992).

\bibitem{Fria94} J.L. Friar and S.A. Coon, {\sl Phys. Rev. C} {\bf 49},
1272 (1994).

\bibitem{Rocha94} C.A. da Rocha and M.R. Robilotta, {\sl Phys. Rev. C}
{\bf 49}, 1818 (1994).

\bibitem{Birse94} M.C. Birse, {\sl Phys. Rev. C} {\bf 49}, 2212 (1994).

\bibitem{Ordo94}
C. Ord\'o\~nez, L. Ray, and U. Van Kolck, {\sl Phys. Rev. Lett.} {\bf 72}, 1982
(1994).

\bibitem{RR95}C.A. da Rocha and M.R. Robilotta, {\sl Phys. Rev. C} 
{\bf 52}, 531 (1995)

\bibitem{CWZ69}
S. Coleman, J. Wess, and B. Zumino, {\sl Phys. Rev.} {\bf 177}, 2239 (1969);
C.G. Callan, S. Coleman, J. Wess, and B. Zumino, {\sl Phys. Rev.} {\bf 177},
2247 (1969).

\bibitem{Rob95} M.R. Robilotta, {\sl Nucl. Phys. A} {\bf 595}, 171 (1995).

\bibitem{BKU95}
V. Bernard, N. Kaiser, and Ulf-G. Meissner, {\sl Int. J. Mod. Phys. E}
{\bf 4}, 193 (1995).

\bibitem{Cal63}
F. Calogero, {\sl Nuovo Cimento} {\bf 27}, 261 (1963); 
{\sl Variable Phase Approach
to Potential Scattering}, (Academic Press, New York, 1967).

\bibitem{BR94}
J.L. Ballot and M.R. Robilotta, {\sl J. Phys. G} {\bf 20}, 1595 (1994).

\bibitem{RW78}
M.R. Robilotta and C. Wilkin, {\sl J. Phys. G} {\bf 4}, L115 (1978).

\bibitem{HJS72} G. H\"ohler, H.P. Jacob, and R. Strauss, {\sl Nucl. Phys. B}
{\bf 39}, 237 (1972).

\bibitem{KP80} R. Kock and E. Pietarinen, {\sl Nucl. Phys. A} {\bf 336},
331 (1980).

\bibitem{HOH83} G. H\"ohler, {\sl Landolt-B\"ornstein Numerical Data and 
Functional Relationships in Science and Technology}, group I, Vol.9,
subvol. B, part 2, Ed. H. Schopper, (1983).

\bibitem{TE78} R. Tarrach and M. Ericson, {\sl Nucl. Phys. A} {\bf 294}, 417
(1978).

\bibitem{Rocha96} M.R. Robilotta and C.A. da Rocha, to be published.
\end{references}
\end{document}